\documentclass[aps,pre,10pt,twocolumn]{revtex4-1}
\usepackage{graphicx}
\usepackage{epstopdf}
\usepackage{amsmath}
\pdfoutput=1
\usepackage{amssymb}
\usepackage[utf8]{inputenc}
\usepackage{graphicx,float,hyperref}
\usepackage{natbib}
\newcommand{\be}{\begin{equation}}
\newcommand{\bea}{\begin{eqnarray}}
\newcommand{\bc}{\begin{center}}            
\newcommand{\ee}{\end{equation}}
\newcommand{\eea}{\end{eqnarray}}
\newcommand{\ec}{\end{center}}

\newcommand{\baa}{\begin{eqnarray*}}
\newcommand{\eaa}{\end{eqnarray*}}
\begin{document}
\title{Optimal operation of a three-level quantum heat engine and universal nature of efficiency}
\author{Varinder Singh}
\email{vsingh@ku.edu.tr}
\affiliation{Department of Physics, Ko\c{c} University, Sar\i{}yer,  Istanbul, 34450, Turkey.} 
\begin{abstract}
We present a detailed study of a three-level quantum heat engine operating at maximum efficient 
power function, a trade-off objective function defined by the product of the efficiency and power output of 
the engine. First, for near equilibrium  
conditions, we find general expression for the efficiency 
and establish universal nature of efficiency at maximum power and maximum efficient power. Then in the high 
temperature limit, optimizing with respect to one parameter while constraining the other one, we obtain the lower and upper 
bounds on the efficiency for both strong as well as weak matter-field coupling conditions. Except for the weak matter-field coupling condition, the obtained bounds on the efficiency exactly match with the bounds already known
for some models of classical heat engines. Further for weak matter-field coupling, we derive some new bounds on
the the efficiency of the the engine which lie beyond the range covered by bounds obtained for strong matter-field coupling.
We conclude by comparing the performance of our three-level quantum heat engine in maximum power and maximum efficient power regimes and show that the engine operating at maximum efficient power produces at least $88.89\%$ of the maximum power output
while considerably reducing the power loss due to entropy production.

\end{abstract} 
%\keywords{quantum brownian motion}
\maketitle
\section{Introduction}
The study of quantum heat engines (QHEs) started with the seminal work of Scovil and Schulz-DuBois (SSD) \cite{Scovil1959}. 
In their work, they investigated the thermodynamics of a three-level maser and showed that its limiting efficiency is 
given by Carnot efficiency \cite{Scovil1967}. Since then, three level systems have been employed to study various models of quantum heat
engines (refrigerators) \cite{Scovil1959B,Geva1994,Geva1996,Geva2002,Scully2001,ScullyAgarwal2003,Linke2005,
BoukobzaTannor2006A,BoukobzaTannor2006B,BoukobzaTannor2007,Scully2011,HarbolaEPL,Harbola2013,Uzdin2015,Harris2016,
KimAgarwal,Cleuren2012,Ghosh2017,Ghosh2018,Dorfman2018,VJ2019,Me6} and quantum absorption refrigerators \cite{Linden2010,Levy2012,Alonso2013,Bijay2017,Segal2018,Scarani2019,Mitchison2016,Brunner2015}. 

Here, we specifically mention the work of Geva and Kosloff \cite{Geva1994,Geva1996,Geva2002} on 
three-level amplifier. They studied the SSD engine in the spirit of finite-time 
thermodynamics using  Alicki's definition of heat and work \cite{Alicki1979}, and optimized its 
performance with respect to  different control parameters. 
They showed that in the presence of external electromagnetic field, one has to incorporate the effect 
of the field on the dissipation superoperators in order to satisfy the second law of thermodynamics. Going one
step further, Tannor and Boukobza formulated a new way of partitioning energy into heat and work \cite{BoukobzaTannor2006A,BoukobzaTannor2006B,BoukobzaTannor2007}. They applied
their formulation to a three-level system simultaneously coupled to two thermal baths at different temperatures and
to a single mode of classical electromagnetic field, and showed that the second law of thermodynamics is always satisfied
without incorporating the effect of the field on the dissipators \cite{BoukobzaTannor2007}. 
Recently, their formalism has been used to study the phenomenon of noise-induced coherence \cite{Dorfman2018}
and quantum synchronization \cite{JaseemSai2020} in nanoscale engines .

In this work, we use Tannor and Boukobza's formalism to analyze the optimal performance of the SSD engine and set up its
correspondence with some classical models of heat engines. At optimal performance, QHEs operating at finite power,
show remarkable similarity to classical macroscopic heat engines. For instance, in high-temperature limit, many models 
of QHEs \cite{Kosloff1984,Geva1992,GevaKosloff1992,LinChen2003,Lutz2012,Deffner2018,Dorfman2018} operate at Curzon-Ahlborn (CA) efficiency, a well known result in the field of finite-time thermodynamics
\cite{Berry1984,Salamon2001,Andresen2011}, first
obtained for a macroscopic model of heat engine known as endoreversible engine \cite{CA1975,Rubin1979}. Similarly, in the low-dissipation 
regime \cite{Esposito2010}, the behavior of quantum and classical heat engines are quite similar 
\cite{Esposito,VascoCavina}.

One another feature common in the operation of classical and QHEs is universal nature of efficiency
\cite{Lindenberg2009}. Many models of classical and QHEs show universality of efficiency at maximum power (EMP)
upto quadratic order in $\eta_C$, i.e., $\eta_{MP}=\eta_C/2+\eta_C^2/8+\mathcal{O}(\eta_C^3)$. Van den Broeck proved
that in the linear response regime, $\eta_C/2$ is universal for tight-coupling heat engines \citep{Broeck2005}. 
Further, Esposito and coauthors established the universality of the second term $\eta_C^2/8$ by invoking the symmetry of
Onsager coefficients on the nonlinear level \cite{Lindenberg2009}. 

The universal features of efficiency are not unique to the EMP, two other optimization 
functions: Omega ($\Omega$) function (or ecological function) \cite{Hernandez2001,ABrown1991} and 
efficient power (EP) function \cite{Stucki,YanChen1996}, $P_\eta=\eta P$ (product of the efficiency  and power of the engine), 
also exhibit this behavior \citep{Zhang2016,Zhang2017}. Here, we will discuss universal character 
of efficiency at maximum efficient power (EMEP) only. The formal proof of universality of EMEP was established in Ref. \cite{Zhang2017}. 
It was shown that the first two universal terms are $2\eta_C/3$ and $2\eta_C^2/27$.

%
%In this work, we study the optimal performance of
%a three-level SSD engine and establish its correspondence with  some classical models of heat engine.

In this paper, we study the optimal performance of the SSD engine operating at maximum efficient power (MEP),
a trade-off optimization function representing a trade-off between the power output and efficiency of a heat engine,
in different operational regimes and compare its performance with the engine operating at maximum power (MP). 
The study of such objective function is important from the environmental and ecological point of view. It is already 
known that engines operating at maximum power regime also waste a lot of power due to large entropy production
\cite{devosbook,Chen2001}. 
Therefore, rather than operating in MP regime, the real  heat engines should operate near MP
regime, where they produce slightly smaller power output with appreciable larger efficiency, which makes 
them cost effective too \cite{Chen2001}.
The EP function was introduced by Stucki \cite{Stucki} in the context of biochemical energy conversion process. Extending Stucki's idea, 
Yan and Chen (YC) 
treated EP function as their objective function to investigate the performance of an endoreversible heat engine
\cite{YanChen1996}. Recently,
EP function has attracted considerable interest and have been employed to study the energy
conversion process in low-dissipation
heat engines \cite{VJ2018,Holubec2015}, thermionic generators \cite{ChenDing}, biological systems \cite{Stucki,ABrown2008},
chemical reactions \cite{Chimal2019,Sanchez2016}, Feynman's ratchet and pawl model \cite{Me5} and in a 
quantum Otto engine \cite{Deffner2020}.

The paper is organized as follows. Sec. II, we discuss  the model of SSD engine. In Sec. III, we obtain
analytic expression for the efficiency of the
SSD engine operating near equilibrium conditions, and show universality of the EMP and EMEP. In Subsecs. IV A and IV B, 
we optimize engine's performance, 
operating in two different operational regimes (strong and weak matter-field coupling regimes), 
with respect to one parameter only, and obtain the lower  
and upper bounds on the EMEP for each case.  Subsec. IV C is devoted to the discussion of universality 
of efficiency for one parameter optimization scheme under the effect of some symmetric constraints imposed
on the control parameters of the engine. In Secs. V and VI, we we compare the performance of the SSD engine 
operating at MEP to the engine operating at MP. We conclude in section VII.
\section{Model of Three Level Quantum Laser Heat Engine}
SSD engine \cite{Scovil1959} is one of the simplest QHEs. Using the concept of 
stimulated emission in a population inverted medium, it converts the incoherent thermal energy of 
heat reservoirs to a coherent laser output. The model consists of a three-level system simultaneously 
coupled to two thermal reservoirs at temperatures $T_h$ and $T_c$ ($T_c<T_h$), and to a single 
mode classical electromagnetic field (see Fig. \ref{SSDmodel}). The hot reservoir at temperature $T_h$ induces the 
transition between the ground state $\vert g\rangle$ and the upper state $\vert 1\rangle$, whereas 
the transition between the middle state 
$\vert 0\rangle$ and the ground state $\vert g\rangle$ is  constantly de-excited by the cold reservoir at 
temperature $T_c$. For power output mechanism, states $\vert 0\rangle$ and
$\vert 1\rangle$ are coupled to a classical single mode field. The bare Hamiltonian of the three-level system is given by: 
$H_0=\hbar \sum \omega_k \vert k\rangle\langle k\vert$ where the sum runs over all three states and 
$\omega_k$'s represent the corresponding atomic frequencies.
Under the rotating wave approximation, the following semiclassical Hamiltonian  describes the interaction of 
the system with the classical  field of frequency $\omega$:
$V(t)=\hbar \lambda (e^{i\omega t} \vert 1\rangle\langle 0\vert + e^{-i\omega t} \vert 0\rangle\langle 1\vert)$; 
$\lambda$ is the field-matter coupling constant. The reduced dynamics of the matter-field system under the effect of 
the heat reservoirs is described by the following form of Lindblad master equation:
\begin{figure}   
 \begin{center}
\includegraphics[width=8.6cm]{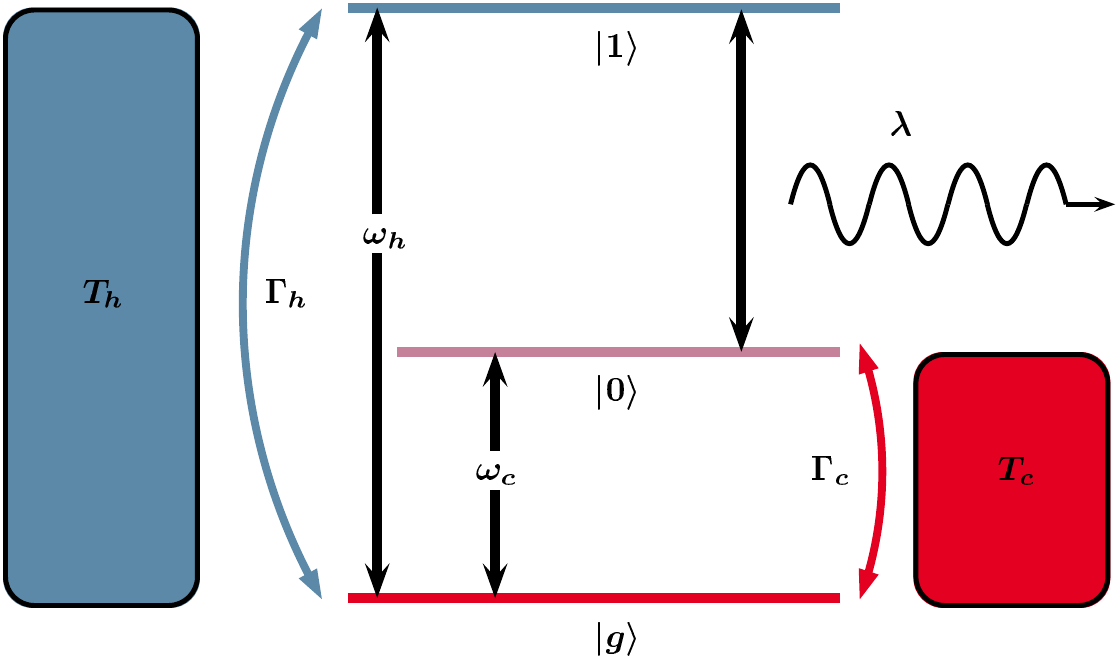}
 \end{center}
\caption{ Model of the SSD engine simultaneously coupled to two thermal reservoirs at temperatures $T_c$ and $T_h$ with
coupling constants $\Gamma_c$ $\Gamma_h$, respectively. The interaction of the system with a classical single-mode field
is represented by $\lambda$, the matter-field coupling constant.}
\label{SSDmodel}
\end{figure}
\begin{equation}
\dot{\rho} = -\frac{i}{\hbar} [H_0+V(t),\rho] + \mathcal{L}_{h}[\rho] + \mathcal{L}_{c}[\rho],
\end{equation}
where $\mathcal{L}_{h}$ and $\mathcal{L}_{c}$ are the dissipation Lindblad superoperator 
describing the  interaction of the system with the hot and cold reservoirs, respectively:
\begin{eqnarray}
\mathcal{L}_h[\rho] &=& \Gamma_h(n_h+1)(2\vert g\rangle\langle g\vert \rho_{11} - \vert 1\rangle\langle 1\vert\rho
- 
\rho\vert 1\rangle\langle 1\vert) \nonumber
\\
&& +\Gamma_h n_h (2\vert 1\rangle\langle 1\vert\rho_{gg}-\vert g\rangle\langle g\vert\rho
-
\rho\vert g\rangle\langle g\vert),\label{dissipator1}
\end{eqnarray} 
\begin{eqnarray}
\mathcal{L}_c[\rho] &=& \Gamma_c(n_c+1)(2\vert g\rangle\langle g\vert \rho_{00} - \vert 0\rangle\langle 0\vert\rho
- 
\rho\vert 0\rangle\langle 0\vert) \nonumber
\\
&& +\Gamma_c n_c (2\vert 0\rangle\langle 0\vert\rho_{gg}-\vert g\rangle\langle g\vert\rho
-
\rho\vert g\rangle\langle g\vert)\label{dissipator2},
\end{eqnarray}
where $\Gamma_c$ and $\Gamma_h$ are Weisskopf-Wigner decay constants, and 
$n_{h(c)}= 1/(\exp[\hbar\omega_{h(c)}/k_B T_{h(c)}]-1)$ is average number of photons in the mode of frequency
$\omega_{h(c)}$
in hot (cold) reservoir satisfying the relations $\omega_c=\omega_0-\omega_g$, $\omega_h=\omega_1-\omega_g$.

In order to solve the density matrix equations, it is convenient to transform to a rotating frame
in which semiclassical interaction Hamiltonian and the steady-state density matrix $\rho_R$
become time-independent \cite{BoukobzaTannor2007}. Defining $\bar{H}=\hbar (\omega_g \vert g\rangle\langle g\vert 
+ \frac{\omega}{2} \vert 1\rangle\langle 1\vert - \frac{\omega}{2} \vert 0\rangle\langle 0\vert ) $, 
an arbitrary operator $B$ in the rotating frame is given by $B_R=e^{i\bar{H}t/\hbar}Be^{-i\bar{H}t/\hbar}$. 
It can be seen that superoperators $\mathcal{L}_c[\rho]$ and $\mathcal{L}_h[\rho]$ remain unchanged under this transformation.
Finally, time evolution of the system density operator in this rotating frame is given by:
\begin{equation}
\dot{\rho_R} = -\frac{i}{\hbar}[H_0-\bar{H}+V_R,\rho_R] + \mathcal{L}_h[\rho_R] + \mathcal{L}_c[\rho_R]\label{dm1}
\end{equation}
where $V_R=\hbar\lambda(\vert 1\rangle\langle 0\vert + \vert 0\rangle\langle 1\vert)$.

For a weak system-bath coupling, the heat flux, output power and efficiency of the SSD engine can be defined, 
using the formalism of Ref. \cite{BoukobzaTannor2007}, as follows:
\begin{eqnarray}
\dot{Q_h} &=&  {\rm Tr}(\mathcal{L}_h[\rho_R]H_0), \label{heat1} \\
P &=& \frac{i}{\hbar} {\rm Tr}([H_0,V_R]\rho_R), \label{power1} \\
\eta &=& \frac{P}{\dot{Q_h}}. \label{efficiencycrude}
\end{eqnarray}
Here, we have used the sign convention in which all three energy fluxes: heat flux extracted from the hot bath,  
heat flux rejected to the cold bath and
the power output  are positive.
Substituting the expressions for $V_R$, $H_0$ and $\mathcal{L}_h[\rho_R]$, and calculating the traces appearing in 
Eqs. (\ref{heat1}) and (\ref{power1})[see Appendix A], the heat flux and power output can be written as:
\begin{equation}
\dot{Q_h}=  i\hbar\lambda\omega_h (\rho_{01}-\rho_{10}),\label{heat2}
\end{equation}
\begin{eqnarray}
P &=& i\hbar\lambda(\omega_1-\omega_0)(\rho_{01}-\rho_{10}), \nonumber
\\
&=& i\hbar\lambda(\omega_h-\omega_c)(\rho_{01}-\rho_{10}), \label{power2} 
\end{eqnarray}
where $\rho_{01} = \langle 0\vert\rho_R\vert 1\rangle$ and $\rho_{10} = \langle 1\vert \rho_R\vert 0\rangle$.
Using Eqs. (\ref{heat2}) and (\ref{power2}) in Eq. (\ref{efficiencycrude}), the efficiency of the engine is given by
\begin{equation}
\eta = 1 - \frac{\omega_c}{\omega_h}. \label{efficiency}
\end{equation}
The positive power production condition [see Eq.(A11)] implies that $\omega_c/\omega_h\geq T_c/T_h$, which in turn implies that
$\eta \leq \eta_C$.
\section{Universal nature of the efficiency}
In this section, we will explicitly show the universal nature of  both  EMP and EMEP. 
The expressions for the power output and EP are derived in Appendix A [Eqs. (A11) and (A12)]. Optimization of these equations with respect to control parameters $\omega_h$ and $\omega_c$ yields very complex equations, which 
cannot be solved analytically under the general conditions. However, close to the equilibrium, they  can be solved 
to give analytic expression for the efficiency upto second order term in $\eta_C$, which is sufficient for our purpose
as we want to focus only on the universal nature of the EMP and EMEP.

As mentioned in the Introduction, the appearance of the first two universal terms in the Taylor series of the EMP was first proven by Esposito and coauthors for tight-coupling heat engines possessing a left-right symmetry in the system.
We briefly outline the algorithm followed in
Ref. \cite{Lindenberg2009}. 
The following formalism is valid for the engines obeying tight-coupling condition between the energy flux 
$I_E$ and matter flux $I$:
\begin{equation}
I_E (x,y) = \epsilon I(x,y), \label{fluxmatter}
\end{equation}
where $x$ and $y$ are dimensionless scaled energies (to be explained later). The above equation implies 
that the energy is exported by the particles of a given energy $\epsilon$. The general formula for the EMP is given by
%
%By optimizing the power of the engine with respect to thermodynamic force, $\mathcal{F}\equiv x-y$, and $x$,
%%
%the general formula for the EMP is obtained as
%
\begin{equation}
\eta = \frac{\eta_C}{2} + \left( 1 + \frac{M}{\partial_{x}L} \right)\frac{\eta_C^2}{4} + \mathcal{O}(\eta_C^3),
\label{effas}
\end{equation}
where
$L=-I'_{1}(x,x)$ and $M=I''_{11}(x,x)/2$. Further, for the systems possessing a left-right symmetry, the inversion of
flux,
\begin{equation}
I(x,y) = -I(y,x),  \label{fluxreversal}
\end{equation}
leads to the condition $2M=-\partial_x L$, which reduces the second term in Eq. (\ref{effas}) to $\eta_C^2/8$, thus
establishing the universality of the coefficient 1/8 under the symmetry specified by Eq. (\ref{fluxreversal}).

In order to inquire the universal character of efficiency in the SSD model, we have to first identify the flux term. 
In our model, energy is transported from hot to cold reservoir by the flux of photons. Comparing Eq. (A11) with Eq. (\ref{fluxmatter}), we identify $I$ as follows
\begin{widetext}
\begin{equation}
I(x,y)=\frac{2 \lambda ^2 \Gamma _c \Gamma _h \left(e^x-e^y\right)}{\lambda ^2 \left[\left(e^x+2\right) \left(e^y-1\right) \Gamma _c+\left(e^x-1\right) \left(e^y+2\right) \Gamma _h\right]+\Gamma _c \Gamma _h \left(e^{x+y}+e^x+e^y\right) \left(\frac{e^x \Gamma _c}{e^x-1}+\frac{e^y \Gamma _h}{e^y-1}\right)}. \label{fluxSSD}
\end{equation}
In the above equation, we have put $x\equiv \hbar\omega_c/k_B T_c$ and $y\equiv \hbar\omega_h/k_B T_h$, and used the
expressions $n_h=1/(e^y-1)$ and $n_c=1/(e^x-1)$. By inspecting
Eq. (\ref{fluxSSD}), we can see that symmetry criterion (\ref{fluxreversal}) is satisfied for $\Gamma_h=\Gamma_c$.
Under this condition, we should observe the universality of efficiency for the SSD model. We confirm this observation
by explicit calculating the form of efficiency in Eq. (\ref{effas}). By evaluating expressions for $L$, $M$ and $\partial_x L$
for the current $I$ given in Eq. (\ref{fluxSSD}), and substituting in Eq. (\ref{effas}), we find
\begin{eqnarray}
\eta &=& \frac{\eta_C}{2} + \frac{\eta_C^2}{4}\left[\frac{\left(\left(e^{\alpha }+1\right) \Gamma _c \left(\left(e^{\alpha }-1\right)^2 \left(e^{\alpha }+2\right) \lambda ^2+e^{2 \alpha } \left(e^{\alpha }-4\right) \Gamma _h^2\right)+e^{3 \alpha } \left(e^{\alpha }-1\right) \Gamma _c^2 \Gamma _h+\left(e^{\alpha }-2\right) \left(e^{\alpha }-1\right)^3 \lambda ^2 \Gamma _h\right)}{ 2\left(\Gamma _c+\Gamma _h\right) \left(\left(e^{\alpha }-1\right)^2 \left(e^{2 \alpha }+2\right) \lambda ^2+e^{2 \alpha } \left(e^{\alpha } \left(e^{\alpha }-2\right)-2\right) \Gamma _c \Gamma _h\right)}\right]\nonumber
\\
&&+ \mathcal{O}(\eta_C^3), \label{efffinal}
\end{eqnarray}
where $\alpha$ is the solution of the transcendental equation \footnote{The transcendental equation is obtained by
substituting the expression for $L$ and $\partial_x L$ in equation $x=-2L/\partial_x L$ (see Ref. \cite{Lindenberg2009}).},
\begin{equation}
\alpha  \left(-\frac{1}{e^{\alpha }+2}+\frac{1}{2-2 e^{\alpha }}+\frac{\left(e^{\alpha }-1\right) \lambda ^2}{\left(e^{\alpha }-1\right)^2 \lambda ^2+e^{2 \alpha } \Gamma _c \Gamma _h}+\frac{1}{2}\right)=1, 
\end{equation}
\end{widetext}
which can be solved by specifying the numerical values of $\lambda$, $\Gamma_h$ and $\Gamma_c$.
For $\lambda^2=\Gamma_h\Gamma_c$, solution of above equation yields $\alpha=2.9327$. For the symmetric dissipation,
$\Gamma_c=\Gamma_h$, the term inside the square bracket in Eq. (\ref{efffinal}) becomes equal to 1/2, yielding the 
coefficient of second term as 1/8, and hence proving our assertion.
\subsection{Universality of the EMEP}
The general expression for the EMEP analogous to Eq. (\ref{effas}) is \cite{Zhang2017}
\begin{equation}
\eta = \frac{2\eta_C}{3} + \left(1 + \frac{M}{\partial_x L}\right)\frac{4\eta_C^2}{27} + \mathcal{O}(\eta_C^3).
\label{effas2}
\end{equation}
Comparing Eqs. (\ref{effas}) and (\ref{effas2}), we can see that in order to obtain the explicit
expression for the EMEP, we just have to replace  $\eta_C^2/4$ by
$4\eta_C^2/27$ in Eq. (\ref{efffinal}); everything else remains the same. Since, for $\Gamma_h=\Gamma_c$, the 
term inside square bracket in Eq. (\ref{efffinal}) is 1/2, we get second term as $2\eta_C^2/27$. 

From the above procedure, we can conclude that if universal nature of the EMP can be established for a model
under consideration, the universality of the EMEP automatically follows. The universal character of the  EMEP
has already been established for the low-dissipation \cite{VJ2018}, endoreversible \cite{Yilmaz,YanChen1996}   
and nonlinear irreversible \cite{Zhang2017} models of classical heat engine. Ours is the first
study of a QHE in which universality of the EMEP is explored and explicitly shown.  

\subsection{Global optimization of efficient power function in low-temperature limit }
\begin{figure} [ht]
 \begin{center} 
\includegraphics[width=8.6cm]{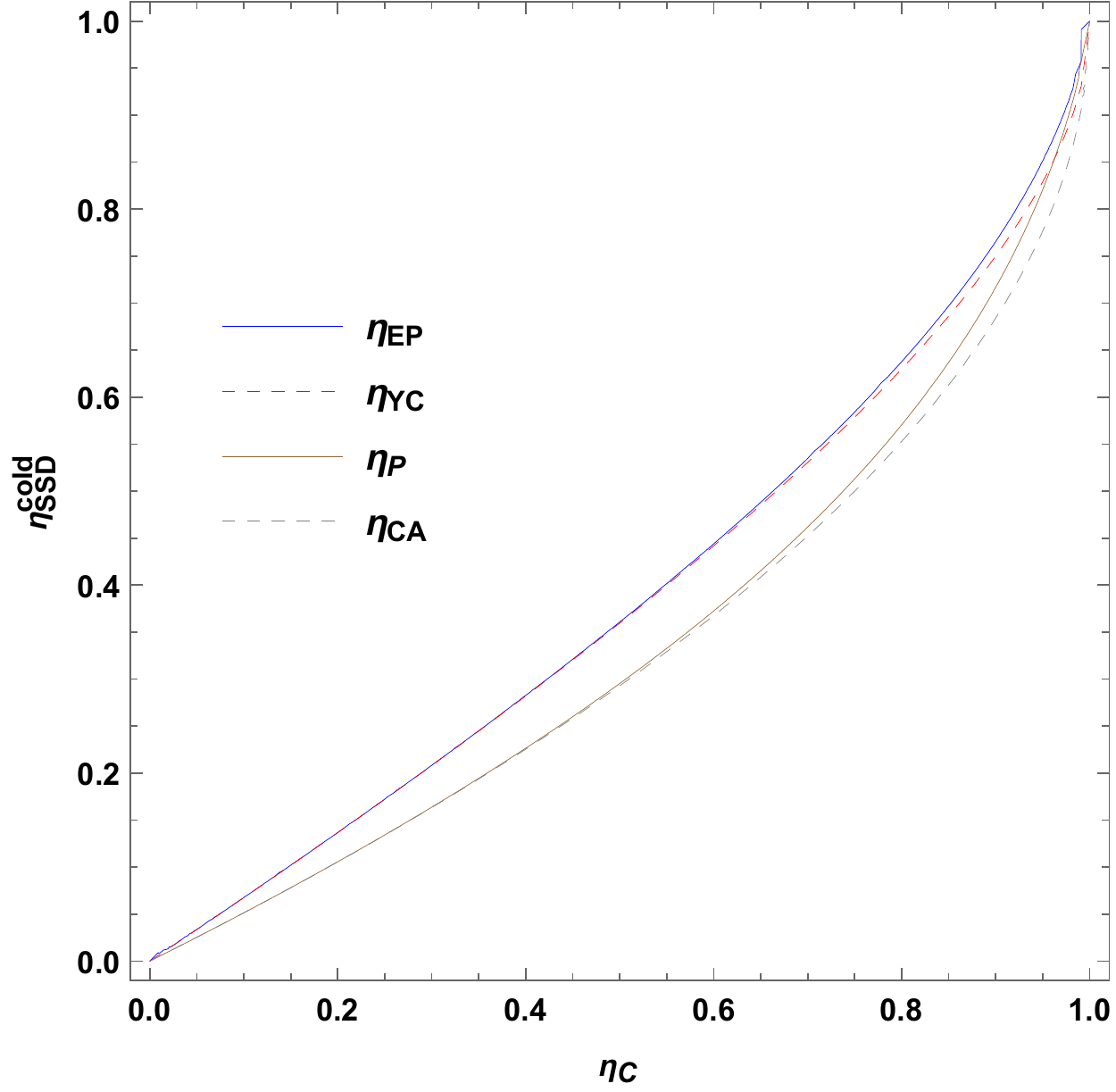}
 \end{center}
\caption{Comparison of the EMEP (EMP) of the SSD engine operating in  low-temperature limit with $\eta_{YC}$ ($\eta_{CA}$). The EMEP (EMP) of the SSD engine is higher than that of low-dissipation
and endoreversible engines which operate at  $\eta_{YC}$ ($\eta_{CA}$).}
\label{SSDcold}
\end{figure}
Now, we study the operation of the SSD engine in the low-temperature regime. In the low-temperature limit, we assume $\hbar\omega_{c,h}\gg k_B T_{c,h}$,
such that $n_{c,h} \simeq e^{-\hbar\omega_{c,h}/k_B T_{c,h}}\gg 1$. The EP function [Eq. (\ref{eco5})] in this case is given by
\begin{equation}
P_\eta = \frac{2\hbar\lambda^2 \Gamma_c\Gamma_h(n_h-n_c)(\omega_h-\omega_c)^2}
{
\omega_h(\Gamma_c + \Gamma_h) (\lambda^2 + \Gamma_c\Gamma_h)} \label{epcold}.
\end{equation}
In our previous work \cite{VJ2019}, we have proven the equivalence of the SSD engine operating in the low-temperature
limit to Feynman's ratchet 
and pawl engine \cite{Feynman,Tu2008,VarinderJohal}, a classical heat engine based on the principle of Brownian fluctuations. Hence, the analysis of this
section is also valid for Feynman's model. 
Maximization of Eq. (\ref{epcold}) with respect to  $\omega_h$ and $\omega_c$, and a little
simplification yields the following equations:
\begin{equation}
 e^{\hbar\omega_h/k_B T_h-\hbar\omega_c/k_B T_c} = 1 - \frac{\hbar\omega_h(\omega_h-\omega_c)}
 {k_B T_h(\omega_h+\omega_c)}, \label{D1}
\end{equation}
\begin{equation}
 e^{\hbar\omega_h/k_B T_h-\hbar\omega_c/k_B T_c} =\frac{2k_B T_c}{\hbar(\omega_h -\omega_c) + 2k_B T_c}. \label{D2}
\end{equation}
It is not possible to obtain analytic solution of these two equations for $\omega_h$ and $\omega_c$. However combining Eqs.
(\ref{efficiency}), (\ref{D1}) and (\ref{D2}), and writing in terms of $\eta_C=1-T_c/T_h$, we obtain following transcendental equation,
\begin{equation}
\frac{(2\eta_C-\eta)(\eta-\eta_C)}{\eta(1-\eta_C)}
=
\ln \left[ \frac{2(1-\eta_C)}{2-\eta}\right]. \label{trans}
\end{equation}
%
%where $\eta_C=1-T_c/T_h$ is Carnot efficiency.
%
It is clear from the Eq. (\ref{trans}) that the efficiency does not depend on the system-parameters and 
depends on $\eta_C$ only. We plot Eq. (\ref{trans}) in Fig. \ref{SSDcold} and compare the EMEP (EMP) of the cold
SSD engine with the corresponding EMEP (EMP) of endoreversible or low-dissipation heat engines.

%Since an analytic solution of Eq. () is not available,
%we look for a perturbative solution in terms of $\eta_C$ near equilibrium. This can be done by
%substituting $\eta=b_1\eta_C+b_2\eta_C^2+b_3\eta_C^3+O(\eta_C^4)$ in Eq. (\ref{trans})
%and then expanding the resulting equation in $\eta_C$. The coefficients $b_1$, $b_2$ and $b_3$
%are found recursively by solving order by order in $\eta_C$. The coefficients of first, second and 
%third order in $\eta_C$  are $a_1=2/3$, $a_2=2/27$ and $a_3=11/243$, respectively. So close to the equilibrium, EMEP 
%behaves as follows
%
Near equilibrium, a perturbative solution for the Eq. (\ref{trans}) is available and is given by \cite{Me5}:
\begin{equation}
\eta^{SSD}_{\rm cold} = \frac{2\eta_C}{3} +  \frac{2\eta_C^2}{27} +  \frac{11\eta_C^3}{243}  + \mathcal{O}(\eta_C^4).
\end{equation}
Hence, again in this regime, we are able to show the existence of the first two universal terms 
($2\eta_C/3$ and $2\eta_C^2/27$) for a two-parameter optimization scheme.
\section{Local optimization}
Since for the unconstrained regime, general solution for the two-parameter optimization of the SSD engine is not available, in the following, 
we will optimize the performance of the engine with respect to one control parameter only while keeping the
other one fixed at a constant value. In the high-temperature limit, it is possible to obtain analytic 
expressions for the EMP and EMEP. In this limit, we put $n_h\approx k_B T_h/\hbar\omega_h$ and 
$n_c\approx k_B T_c/\hbar\omega_c$.
\subsection{High temperature and strong coupling regime}
Assuming the strong matter-field coupling ($\lambda\gg\Gamma_{h,c}$), the expression for EP function
in the high-temperature limit can be written as
\begin{equation}
P_\eta = \frac{2\hbar\Gamma_h(\omega_h-\omega_c)^2(\omega_c-\tau\omega_h)}{3\omega_c(\omega_c\gamma+\tau\omega_h)}.
\label{EPhot}
\end{equation}
It is important to mention that two parameter optimization of EP function given in Eq. (\ref{EPhot})
is not possible. Such two parameter optimization scheme leads to the trivial solution, 
$\omega_c=\omega_h=0$, which is  not a useful result.
It indicates that a unique maximum of $P_\eta$ with respect
to both $\omega_c$ and $\omega_h$ cannot exist. It can be argued as follows. For the
given values of the bath temperatures and the coupling constants, 
under the scaling $(\omega_c, \omega_h) \to (\beta \omega_c, \beta \omega_h)$, where
$\beta$ is a certain positive number, the EP function also scales as $P_\eta \to \beta P_\eta$.
Hence, there cannot exist a unique optimal solution $({\omega}^*_h$, ${\omega}^*_c)$  
that yields a unique maximum for EP function. The same is also true for the optimization of power output
of the engine with respect to $\omega_c$ and $\omega_h$.
Therefore we will perform  optimization with respect to one parameter only while keeping the other one 
fixed. First we keep $\omega_h$ fixed, then optimizing EP [Eq.(\ref{EPhot})] with respect to
$\omega_c$, i.e., setting $\partial P_\eta/\partial\omega_c=0$, we evaluate EMEP as
\begin{equation}
\eta^{P\eta}_{\omega_h} = 1-\frac{(\gamma -3) \tau +\sqrt{\gamma +1} \sqrt{\tau } \sqrt{\gamma  (\tau +8)+9 \tau }}{4 \gamma },
\label{epgeneral}
\end{equation}
where $\gamma=\Gamma_h/\Gamma_c$. For a given value of $\tau$, $\eta^{P\eta}_{\omega_h}$ is a monotonically increasing function of $\gamma$. Hence, we
can obtain the lower and upper bounds of EMEP by setting $\gamma\rightarrow 0$ and $\gamma\rightarrow\infty$, respectively.
Writing in terms of $\eta_C = 1-\tau$, we have 
\begin{equation}
 \frac{2\eta_C}{3} \leq \eta^{P\eta}_{\omega_h}  \leq 1-\frac{1}{4} (1-\eta_C) \left(1+\sqrt{1+\frac{8}{1-\eta_C}}\right)
\equiv \eta_{_{YC}}. \label{boundsep}
\end{equation}
Recently, $\eta^{P_\eta}_-\equiv 2\eta_C/3$, has also been obtained as the lower bound of the low-dissipation 
heat engines operating at MEP \cite{VJ2018,Holubec2015}.
The upper bound $\eta_{_{YC}}$ obtained here, was first obtained by 
Yan and Chen  for  an endoreversible heat engine \cite{YanChen1996}. Hence, we name it after them. $\eta_{_{YC}}$
can also be obtained for symmetric low-dissipation heat engines \cite{VJ2018}. 

Alternately, we may fix the value of $\omega_c$ and optimize 
EP function with respect to $\omega_h$. In this case, we get following  equation:
%
%\begin{widetext}
\begin{equation}
\frac{\tau ^2 \omega _h^3 + \tau  (2 \gamma +\tau ) \omega _c \omega _h^2 - (\gamma +2 \tau ) \omega _c^2 \omega _h
-
\gamma  \omega _c^3}{\omega_c\gamma+\omega_h\tau}=0. \label{casus}
\end{equation}
%\end{widetext}
%
Due to Casus irreducibilis (see Appendix \ref{casusir}), the roots of the cubic polynomial in numerator  
of Eq. (\ref{casus}) can only be expressed in terms
of complex radicals, although the roots are real actually. Still, Eq. (\ref{casus}) can be solved  
for the limiting cases $\gamma\rightarrow 0$ 
and  $\gamma\rightarrow\infty$. For $\gamma\rightarrow 0$, the EMEP is evaluated at YC value. 
For $\gamma\rightarrow\infty$, we obtain $\eta^{P_\eta}_+=(3-\sqrt{9-8\eta_C})/2$.
Hence, EMEP lies in the range
\begin{equation}
\eta_{_{YC}} \leq \eta \leq \frac{1}{2} \left(3-\sqrt{9-8\eta_C}\right) \equiv \eta^{P_\eta}_+. \label{boundsep2}
\end{equation}
Upper bound  $\eta^{P_\eta}_+$ obtained here also serves as the upper bound of the low-dissipation
model of heat engine \cite{VJ2018,Holubec2015}. The same expression also appears in the optimization
of Feynman's model operating at MEP in high-temperature regime \cite{Me5}.

The corresponding efficiency bounds for the optimization of power output of the SSD engine is given by \cite{Dorfman2018}
\begin{equation}
\eta^P_-\equiv \frac{\eta_C}{2} \leq \eta^P_{\omega_h} \leq 1 - \sqrt{1-\eta_C}\equiv \eta_{CA}, \label{B11}
\end{equation}
\begin{equation}
\eta_{CA} \leq \eta^P_{\omega_c} \leq  \frac{\eta_C}{2-\eta_C} \equiv \eta^P_+. \label{BB}
\end{equation}
Comparing Eqs. (\ref{boundsep}) and (\ref{boundsep2}) with Eqs. (\ref{B11})-(\ref{BB}), we can conclude that the SSD 
engine operating under MEP is
far more efficient than the engine operating at MP (see Fig. 3).
\subsection{Weak coupling in high-temperature regime}
\begin{figure}   
 \begin{center}
\includegraphics[width=8.6cm]{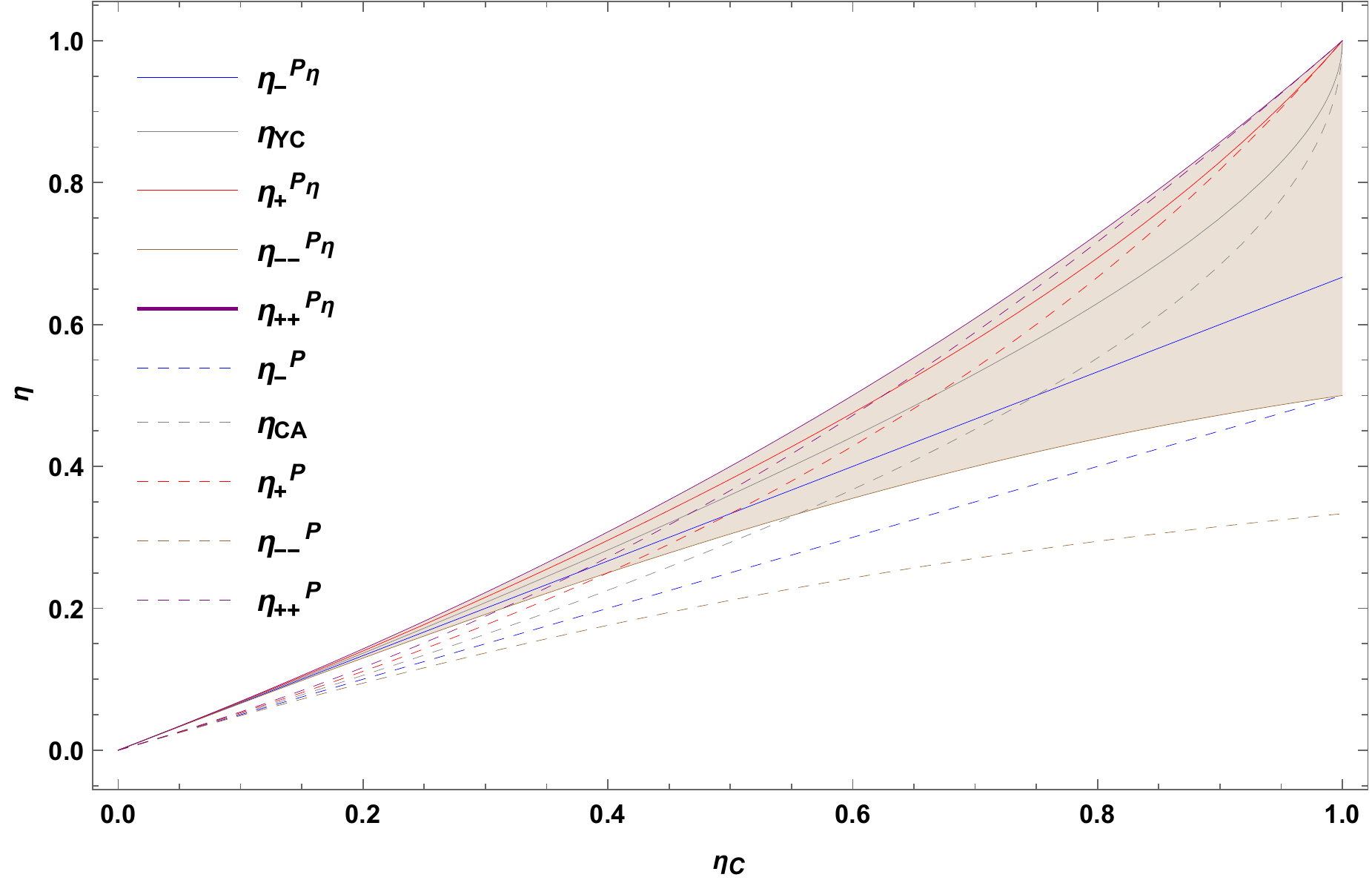}
 \end{center}
\caption{Efficiency $\eta$ versus the Carnot efficiency $\eta_C$ for different operational regimes. The curved under
shaded area represent EMEP. Dashed curves represent corresponding EMP.}
\label{effcurves}
\end{figure}
In addition to the strong matter-field coupling regime ($\lambda\gg\Gamma_{h,c}$), we can also find  analytic 
expressions for the efficiency in weak matter-field coupling regime ($\lambda\ll\Gamma_{h,c}$). 
In the high-temperature limit ($n_{c,h}\gg 1$),  the first two terms in the 
denominator of Eq. (A12) can be ignored. 
Plus we need extreme dissipation conditions, i.e., either  $\Gamma_c\gg\Gamma_h$ ($\gamma\rightarrow 0$) 
or $\Gamma_c\ll\Gamma_h$ ($\gamma\rightarrow\infty$). Hence, for $\gamma\rightarrow 0$ and 
$\gamma\rightarrow\infty$, Eq. (A12) can be approximated by the following two equations, respectively:
\begin{equation}
P_{\eta(\gamma\rightarrow 0)} = \frac{2\hbar\lambda^2 (n_h-n_c) (\omega_h-\omega_c)^2}
{
 3  \omega_h n_h n_c^2 \Gamma_c}, \label{auxi1}
\end{equation}
\begin{equation}
P_{\eta(\gamma\rightarrow\infty)} = \frac{2\hbar\lambda^2 (n_h-n_c) \omega_h-\omega_c)^2}
{
 3  \omega_h n_c n_h^2 \Gamma_h}.   \label{auxi2}
\end{equation}
Optimization of Eqs. (\ref{auxi1}) and (\ref{auxi2}) with respect to $\omega_c$ ($\omega_h$ fixed) yields the following 
bounds on the efficiency
\begin{equation}
\eta^{P_\eta}_{--}\equiv \frac{1}{8} \Big[3(1+\eta_C)-\sqrt{9\eta_C^2-14\eta_C+9}\Big]
\leq \eta^{P'_\eta}_{\omega_h} \leq
 \frac{2\eta_C}{3}. \label{auxi3}
\end{equation}
Note that the above bounds lie below the the parametric area bounded by the efficiency curves given in 
Eq. (\ref{boundsep}) (see Fig. \ref{effcurves}). To the best of our knowledge, these are the new bounds 
which are not previously obtained for any  model of classical or QHE.
Similarly, for the optimization of Eqs. (\ref{auxi1}) and (\ref{auxi2}) with respect to $\omega_h$ ($\omega_c$ fixed),
EMEP lies in the range
\begin{equation}
\frac{1}{2}\left( 3-\sqrt{9-8\eta_C}  \right) \leq  \eta^{P'_\eta}_{\omega_h} \leq \frac{2\eta_C}{3-\eta_C}
\equiv \eta^{P_\eta}_{++}. \label{auxi4}
\end{equation}
Similar to the above case, the efficiency curves in Eq. (\ref{auxi4}) lie above the  parametric area 
bounded by the efficiency curves given in 
Eq. (\ref{boundsep2}) (see Fig. \ref{effcurves}).

The corresponding expressions for the EMP show similar trend, and, are given by
\begin{equation}
\eta^P_{--}\equiv \frac{1}{3} \Big( 1+\eta_C-\sqrt{1-\eta_C+\eta_C^2} \Big) \leq \eta^{P'}_{\omega_h}  \leq \frac{2\eta_C}{3} \label{auxi5},
\end{equation}
\begin{equation}
 \frac{\eta_C}{2-\eta_C} \leq \eta^{P'}_{\omega_c} \leq -1+\eta_C+\sqrt{1-\eta_C+\eta_C^2} \equiv \eta^P_{++}. \label{auxi6}
\end{equation}

We summarize our findings in Table I. As  can be seen from Table I, Taylor's series expansions for different expressions
for the EMP and EMEP show very interesting behavior. For the EMEP, the first universal term $2\eta_C/3$ is present in all 
cases, and the second terms constitute an arithmetic series with common difference $2\eta_C^2/27$. Similarly for the EMP, 
the first universal term $\eta_C/2$ is present in all cases, and the second  term increases by $\eta_C^2/8$
in going from the first case to the last case. Additionally, the third terms in the series expansion of various
forms of the EMP also constitute an arithmetic series with common difference $\eta_C^3/16$.  

\begin{table*}
\caption{Taylor series expansions for the various forms of EMEP and EMP obtained under different operational conditions.}
\renewcommand{\arraystretch}{1.5}
\centering
\begin{tabular}{|c|c|c|c|c|}
\hline
EMEP&EMP\\ \hline
$\eta^{P_\eta}_{--}=\frac{2}{3}\eta_C-\frac{2}{27}\eta_C^2-\frac{14}{243}\eta_C^3+O(\eta_C^4)$ 
&
$\eta^{P}_{--}=\frac{1}{2}\eta_C-\frac{1}{8}\eta_C^2-\frac{1}{16}\eta_C^3+O(\eta_C^4)$ 
\\ \hline
$\eta^{P_\eta}_{-}=\frac{2}{3}\eta_C$ & $\eta^{P}_{-}=\frac{1}{2}\eta_C$ 
\\ \hline
$\eta_{YC}=\frac{2}{3}\eta_C+\frac{2}{27}\eta_C^2+\frac{10}{243}\eta_C^3+O(\eta_C^4)$ 
&
$\eta_{CA}=\frac{1}{2}\eta_C+\frac{1}{8}\eta_C^2+\frac{1}{16}\eta_C^3+O(\eta_C^4)$
\\ \hline
$\eta^{P_\eta}_{+} = \frac{2}{3}\eta_C+\frac{4}{27}\eta_C^2+\frac{16}{243}\eta_C^3+O(\eta_C^4)$  
& 
$\eta^{P}_{+}=\frac{1}{2}\eta_C+\frac{2}{8}\eta_C^2+\frac{2}{16}\eta_C^3+O(\eta_C^4)$  
\\ \hline
$\eta^{P_\eta}_{++}=\frac{2}{3}\eta_C+\frac{6}{27}\eta_C^2+\frac{18}{243}\eta_C^3+O(\eta_C^4)$
& 
$\eta^{P}_{++}=\frac{1}{2}\eta_C+\frac{3}{8}\eta_C^2+\frac{3}{16}\eta_C^3+O(\eta_C^4)$\\ \hline
\end{tabular}\label{tab:eng-maaath}
\end{table*}
\subsection{Universality of efficiency in one-parameter optimization}
Now, we explore the universal nature of efficiency for one-parameter optimization scheme. 
We notice that if we put $\Gamma_c=\Gamma_h$ ($\gamma=1$), in Eq.
(\ref{epgeneral}), the obtained form of the efficiency, 
\begin{eqnarray}
\eta^{P_\eta}_{\omega_h(\gamma=1)}&=& \frac{1}{2}(3-\eta_C-\sqrt{(1-\eta_C)(9-5\eta_C)})\nonumber
\\
&=&
\frac{\eta_C}{3}+\frac{\eta_C^2}{27}+\mathcal{O}(\eta_C^3), 
\end{eqnarray}
does not include the second universal term
$2\eta_C^2/27$ unlike the case of global optimization over the two parameters as shown in section II. 
We attribute this to the nature of optimization scheme. The parametric space available to the control 
variables is different for two different optimization schemes, hence explaining the difference between them. 
However, one can still retain the second order universal term $\eta_C^2/27$ if one imposes
an extra symmetric condition on the constraints of the optimization. The constraints are symmetric
in the sense that under the exchange of the control variables, the constraint equation remains unchanged. 
The physics of such constraints is explored in the Ref. \cite{UzdinEPL}. Here, we want to focus only
on the universal character of the  efficiency under such constraints. 
For instance, if we impose a symmetric
constrain, viz, $\omega_c+\omega_h=k$, optimization of Eq. (\ref{EPhot}) with respect to $\omega_h$ leads to 
the following equation:
\begin{equation}
 x^3 + \frac{-(3-7\tau)}{2(1-\tau )}x^2 + \frac{-2\tau(3+4\tau)}{2(1-\tau)(1+\tau)}x + \frac{\tau(2+3\tau)}
 {2(1-\tau)(1+\tau}=0,
\end{equation}
where we have put $x=\omega_c/k$ for simplicity. The above equation is not solvable in terms of real radicals due to Casus irreducibilis (see Appendix \ref{casusir}). However, the equation can be solved in terms of trigonometric functions \cite{BarnettBook}, 
and the solution is given by
\begin{widetext}
\begin{equation}
x = -\frac{A}{3} + \frac{2}{3}\sqrt{A^2-3B} \,
\cos\left[\frac{1}{3}\arccos\left( -\frac{2A^3-9A B+27C}{2(A^2-3B)^{3/2}}  \right)  \right],
\end{equation}
\end{widetext}
where $A=-(3-7\tau)/2(1-\tau )$, $B=-2\tau(3+4\tau)/2(1-\tau)(1+\tau)$ and $C=\tau(2+3\tau)$. 
Substituting above expression in Eq. $\eta=1-\omega_c/\omega_h=(1-2x)/(1-x)$ (using $\omega_h=k-\omega_c$), and
taking its series expansion, we have 
\begin{equation}
\eta^{P_\eta}_k = \frac{2}{3}\eta_C + \frac{2}{27}\eta_C^2 + \mathcal{O}(\eta_C^3),
\end{equation}
which clearly shows the presence of the first two universal terms.
In the similar manner, for  another symmetric constraint $\omega_c\,\omega_h=k'$, the expression for 
efficiency, $\eta = 1-k'/\omega_h^2$, turns out to be 
\begin{eqnarray}
\eta^{P_\eta}_{k'} &=& 1- \frac{3\tau^2 M^{1/3}}{9\tau^4+36\tau^3+19\tau^2 + M[M-\tau  (3 \tau +4)]} \nonumber
\\
&=& \frac{2}{3}\eta_C + \frac{2}{27}\eta_C^2 + \frac{23}{486}\eta_C^3 + \mathcal{O}(\eta_C^4),
\end{eqnarray}
and we again retain the second universal term $\eta_C^2/27$. Here, $M=\sqrt{k}f(\tau)$, is function of
$\tau$ only.

We can also obtain the first two universal terms ($\eta_C/2$ and $\eta_C^2/8$) in the series
expansion of the EMP for the optimization under symmetric constraints.
Thus for the SSD model, we have shown that in order to establish the universality of efficiency upto 
the quadratic order term in $\eta_C$, we have to impose an additional symmetric condition in addition to the condition $\gamma=1$. 
Similar is also true for the optimization of an ultra hot Otto engine \cite{UzdinEPL} and Feynman's ratchet model \cite{VarinderJohal2018} both of which possess a certain left-right symmetry in the system.

\section{Fractional loss of power at maximum ecological function and maximum power output}
In this section, we make a comparison of the performance of the SSD engine operating at MEP to
that of operating at MP. In both cases, we find the expressions for the fractional 
loss of power due to entropy production, $\dot{S}_{tot}=\dot{Q}_c/T_c-\dot{Q}_h/T_h$.
Power loss due to entropy production is given by: $P_{\rm lost}=T_2\dot{S}_{\rm tot}=\dot{Q}_c-(1-\eta_C)\dot{Q}_h$.
Further using the definitions of power output $P=\dot{Q}_h-\dot{Q}_c$ and efficiency $\eta=P/\dot{Q}_h$, the ratio
of power loss to power output can be derived as:
\begin{equation}
R \equiv \frac{P_{\rm lost}}{P} = \frac{\eta_C}{\eta}-1. \label{ratioplost}
\end{equation}
We calculate the ratio $R$ in four different cases: two for the optimization
of EP with respect to $\omega_c$ ($\omega_h$ fixed) in the extreme dissipation limits when
$\gamma\rightarrow 0$ and $\gamma\rightarrow\infty$, and similar two cases for optimization
with respect to $\omega_h$ ($\omega_c$ fixed). Using Eqs. (\ref{boundsep}) and (\ref{ratioplost}), we have
\begin{widetext}
\begin{equation}
R^{P_\eta}_{\omega_h(\gamma\rightarrow 0)} = \frac{1}{2},\,
R^{P_\eta}_{\omega_h(\gamma\rightarrow\infty)} =
 \frac{1}{4}\left[\sqrt{(9-\eta_C)(1-\eta_C)}-(1-\eta_C)\right]. \label{PL1}
\end{equation}
Similar equations for the optimization of $P_\eta$ with respect to $\omega_h$ for a fixed $\omega_c$ can be obtained
by using Eqs. (\ref{boundsep2}) and (\ref{ratioplost}):
\begin{equation}
R^{P_\eta}_{\omega_c(\gamma\rightarrow 0)} = R^{P_\eta}_{\omega_h(\gamma\rightarrow\infty)}, \,
R^{P_\eta}_{\omega_c(\gamma\rightarrow\infty)} =  \frac{1}{4} \left(\sqrt{9-8 c}-1\right) .\label{PL2}
\end{equation}
\end{widetext}
\begin{figure} [ht]
 \begin{center} 
\includegraphics[width=8.6cm]{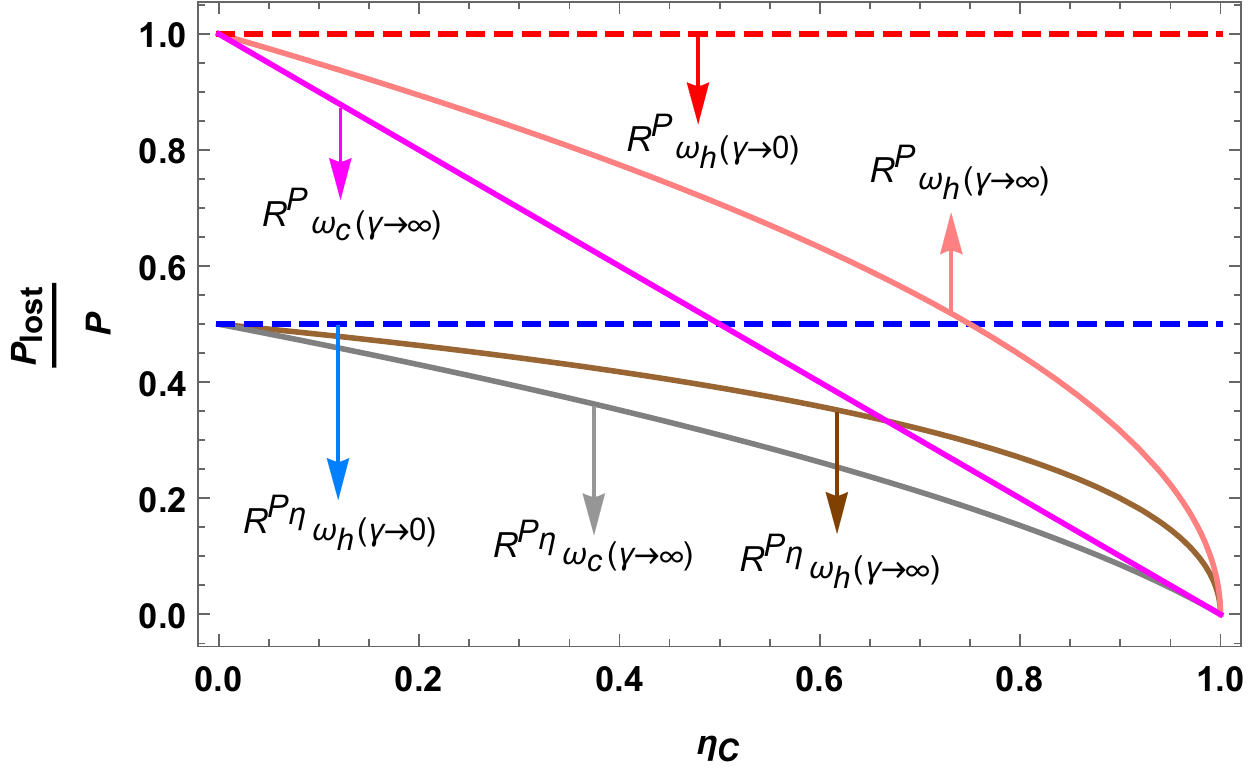}
 \end{center}
\caption{Comparison of the ratios of power lost to useful power for   
 two different optimization functions - EP function and power output. The lower lying curves 
[Eqs. (\ref{PL1}) and (\ref{PL2})] represent the case when EP function is optimized whereas the upper 
lying curves [Eqs. (\ref{PL3}) and (\ref{PL4})] represent the corresponding case for the 
optimization of power output.}
\label{powerlost}
\end{figure}
For near equilibrium conditions ($\eta_C\rightarrow 0$), all  above equations approach the value 1/2  (also see Fig. \ref{powerlost}). The fractional loss of
power is lower for the case with fixed $\omega_c$ than with a fixed $\omega_h$. With increasing Carnot 
efficiency $\eta_C$, fractional loss of power decreases, which is natural as for higher $\eta_C$, 
engine also operates with higher efficiency wasting less fuel. Also note that 
$R^{P_\eta}_{\omega_c(\gamma\rightarrow 0)} = R^{P_\eta}_{\omega_h(\gamma\rightarrow\infty)}$,
as expected, since the corresponding efficiencies are also equal, 
$\eta^{P_\eta}_{\omega_h(\gamma\rightarrow\infty)}=\eta^{P_\eta}_{\omega_c(\gamma\rightarrow 0)}=\eta_{_{YC}}$,
as can be seen from Eqs. (\ref{boundsep}) and (\ref{boundsep2}).
For the SSD engine operating at MP, the ratios of power loss to power output for the different
cases discussed above, are given by \cite{VJ2019}:
\begin{equation}
R^{P}_{\omega_h(\gamma\rightarrow 0)} = 1,\,
R^{P}_{\omega_h(\gamma\rightarrow\infty)} = \sqrt{1-\eta_C}. \label{PL3}
\end{equation}
\begin{equation}
R^{P}_{\omega_c(\gamma\rightarrow 0)} = R^{P}_{\omega_h(\gamma\rightarrow\infty)}, \,
R^{P}_{\omega_c(\gamma\rightarrow\infty)} =  1-\eta_C. \label{PL4}
\end{equation}
As can be seen from Fig. \ref{powerlost}, the curves representing the optimal power case
follow the same trend as noted for the optimal EP. More importantly, for small values of $\eta_C$ (near equilibrium),
the curves (lower set) for optimal EP lie well below the curves (upper set) for optimal power.
We specifically mention the case where 
$R^{P}_{\omega_h(\gamma\rightarrow 0)} = 1$, which implies that in this case,  
power loss is equal to the power output. The corresponding case for the optimal EP presents us with much better scenario.
In this case, $R^{P_{\eta}}_{\omega_h(\gamma\rightarrow 0)} = 1/2$, which implies that loss of power is half of the power output. It indicates that EP function is a good target function to optimize if our preference is fuel
conservation.

\section{Ratio of power at maximum efficient power to maximum power}
\begin{figure}   
 \begin{center}
\includegraphics[width=8.6cm]{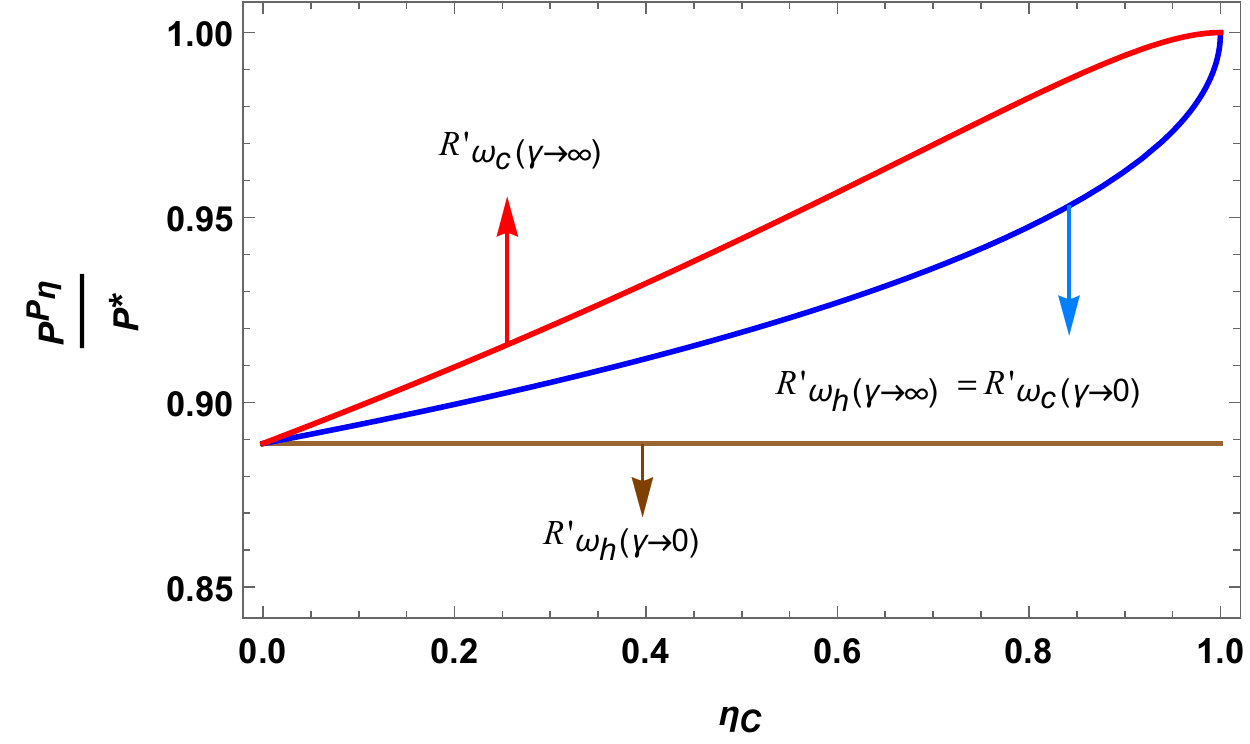}
 \end{center}
\caption{ Comparison of the ratio $R'$ of the power output at MEP 
to the MP [Eqs. (\ref{ratiop1}) and (\ref{ratiop2})].}
\label{ratioPEP}
\end{figure}

Since the fractional loss of power is less when our SSD engine 
operates at MEP as compared to the the case when engine is
operating at MP, it is useful to calculate
the ratio ($R'$) of power at MEP to optimal power.  Dividing Eq. (\ref{peffp}) by Eq. (\ref{optimalp}), and 
taking the limits $\gamma\rightarrow 0$ and $\gamma\rightarrow\infty$, we get following two expressions, respectively:
\begin{widetext}
\begin{equation}
R'_{\omega_h(\gamma\rightarrow 0)} = \frac{8}{9}, 
\quad 
R'_{\omega_h(\gamma\rightarrow\infty)} = \frac{9-5\eta_C-3\sqrt{(1-\eta_C)(9-\eta_C)}}{4(1-\sqrt{1-\eta_C})^2}.
\label{ratiop1}
\end{equation}
Similar equations can be obtained for optimization over $\omega_h$ (fixed $\omega_c$), and
are given by
\begin{equation}
R'_{\omega_c(\gamma\rightarrow 0)} =R'_{\omega_h(\gamma\rightarrow\infty)}, 
\quad
R'_{\omega_c(\gamma\rightarrow 0)} = \frac{(3-\sqrt{9-8\eta_C})(4\eta_C-3+\sqrt{9-8\eta_C})}{4\eta_C^2}.
\label{ratiop2}
\end{equation}
\end{widetext}
For very small temperature differences, i.e., $\eta_C\rightarrow 0$, the ratio $R'=8/9$, which shows that at least
88.89$\%$ of the MP is produced in the MEP regime, which is a considerable amount keeping in mind that 
the power loss in MEP regime is at least 1/2 of the case when engine operates at MP. It is clear
from Fig. \ref{ratioPEP} that ratio $R'$ increases with increasing $\eta_C$, which is expected behavior since the efficiency also increases, while the dissipation decreases.

\section{Concluding Remarks}
We have thoroughly investigated the  performance of the SSD engine operating under the conditions
of MEP and side by side compared its performance with the engine operating at MP. 
First, for close to the equilibrium conditions, we found a analytic solution for the efficiency
of the SSD engine and explicitly showed the universality of the first two terms of both EMP and
EMEP under the  symmetric dissipation ($\gamma=1$) condition. Then, 
we carried out optimization of EP function alternatively with respect to one of the control frequencies 
$\omega_c$ or $\omega_h$ while
keeping the other one fixed at a constant value. In the high-temperature limit, we were able to find lower and upper
bounds on the EMEP  for strong ($\lambda\gg\Gamma_{h,c}$) as well as for weak ($\lambda\ll\Gamma_{h,c}$) 
matter-field coupling conditions. Then, we showed that the obtained form of the EMEP in case of one parameter 
optimization shows universal features of efficiency only in the presence of an extra symmetry imposed on the control 
parameters of the engine. It is important to highlight that except for the weak matter-field 
coupling ($\lambda\ll\Gamma_{h,c}$) condition, the obtained expressions
of the EMEP and EMP in all cases discussed above are same as obtained for different models of 
classical heat engine. Specifically, in weak matter-field coupling regime, we obtained some new bounds on the efficiency
of the SSD engine which lie beyond the  area covered by bounds obtained for strong matter-field coupling.

Finally, we have compared the optimal 
performance of the SSD engine operating at MEP to that of operating at MP. It can be inferred that 
fraction loss of power due to entropy production is appreciably low  in the case of heat engine operating 
at MEP while at the same time it produces at least $88.89\%$ of the MP output. This 
indicates that EP function is a good optimization function  
and real engines should be designed to operate  along the lines of maximizing EP function if our preference
is environment and fuel conservation.
\section*{Acknowledgments}
The author gratefully acknowledges useful and fruitful discussions with Kirandeep Kaur.
\begin{widetext}
\appendix\section{Steady state solution  of density matrix equations}
Here, we provide steps to solve the equations for density operator in the steady state. Substituting the expressions for $H_0$, $\bar{H}$,
$V_0$, and using Eqs. (\ref{dissipator1}) and (\ref{dissipator2}) in Eq. (\ref{dm1}), the time evolution
of the  matrix elements of the density operator are given by following equations:
\begin{eqnarray}
\dot{\rho}_{11} &=& i\lambda (\rho_{10}-\rho_{01}) - 2\Gamma_h[(n_h+1)\rho_{11}-n_h\rho_{gg}],\label{A1} \\
\dot{\rho}_{00} &=& -i\lambda (\rho_{10}-\rho_{01}) - 2\Gamma_c[(n_c+1)\rho_{00}-n_c\rho_{gg}], \\
\dot{\rho}_{10} &=& -[\Gamma_h(n_h+1)+\Gamma_c(n_c+1)]\rho_{10} + i\lambda(\rho_{11}-\rho_{00}), \nonumber
\\
\\
\rho_{11} &= & 1-\rho_{00} - \rho_{gg}, \\
\dot{\rho}_{01} &=& \dot{\rho}_{10}^*. \label{A5}
\end{eqnarray}
Solution of the Eqs. (\ref{A1}) - (\ref{A5}) in the steady state ($\dot{\rho}_{mn}=0$ ($m,n=0,1$)) yields:
\begin{equation}
\rho_{10} = \frac{i\lambda(n_h-n_c)\Gamma_c\Gamma_h}
{
\lambda^2[(1+3n_h)\Gamma_h + (1+3n_c)\Gamma_c] + \Gamma_c\Gamma_h[1+2n_h+n_c(2+3n_h)][(1+n_c)\Gamma_c + (1+n_h)\Gamma_h ] 
},\label{rho10}
\end{equation}
and 
\begin{equation}
\rho_{01} = \rho_{10}^*. \label{rho01}
\end{equation}
Evaluation of the trace in Eq. (\ref{power1}) leads to the following form of output power, 
\begin{equation}
P = i\hbar\lambda(\omega_h-\omega_c)(\rho_{01}-\rho_{10}), \label{power3}
\end{equation}
Similarly calculating the trace in Eq. (\ref{heat1}), heat flux $\dot{Q}_h$ from the hot reservoir can be written as
\begin{equation}
\dot{Q}_h = -\hbar \omega_h (2\Gamma_h[(n_h+1)\rho_{11}-n_h\rho_{gg}]). \label{heat3}
\end{equation}
Employing the steady state condition $\dot{\rho}_{11}=0$ [Eq. (\ref{A1})], Eq. (\ref{heat3}) becomes
\begin{equation}
\dot{Q}_h =  i\hbar \lambda\omega_h(\rho_{01}-\rho_{10}).
\end{equation}
Substituting Eqs. (\ref{rho10}) and (\ref{rho01}) in Eq. (\ref{power3}), we get final expression for the power output.
Since EP function is just power output multiplied by the efficiency, we have following expressions for power
output and EP, respectively:
\begin{equation}
P =    \frac{2\hbar\lambda^2 \Gamma_c\Gamma_h(n_h-n_c) (\omega_h-\omega_c) }
{
\lambda^2[(1+3n_h)\Gamma_h + (1+3n_c)\Gamma_c] + \Gamma_c\Gamma_h[1+2n_h+n_c(2+3n_h)][(1+n_c)\Gamma_c + (1+n_h)\Gamma_h] 
},
\end{equation}
\begin{equation}
P_\eta = \frac{2\hbar\lambda^2 \Gamma_c\Gamma_h(n_h-n_c)(\omega_h-\omega_c)^2}
{
\omega_h\lambda^2[(1+3n_h)\Gamma_h + (1+3n_c)\Gamma_c] + \Gamma_c\Gamma_h[1+2n_h+n_c(2+3n_h)][(1+n_c)\Gamma_c + (1+n_h)\Gamma_h] 
}.\label{eco5}
\end{equation}
\section{Optimization of $P$ and $P_\eta$ in high-temperature and strong-coupling regime}
In the high temperatures limit,  $n_h$ and $n_c$ can be approximated as
\begin{equation}
n_h =\frac{1}{e^{\hbar\omega_h/k_B T_h}-1}\simeq \frac{k_B T_h}{\hbar\omega_h},
\quad
n_c = \frac{1}{e^{\hbar\omega_c/k_B T_c}-1}\simeq \frac{k_B T_c}{\hbar\omega_c}. \label{nc}
\end{equation}
Using Eq. (\ref{nc})  in Eq. (A11) and (A12) and dropping the terms containing $\Gamma_{c,h}$ in 
comparison to $\lambda$, we can write $P$ and $P_\eta$ in terms of  $\gamma=\Gamma_h/\Gamma_c$ and $\tau=T_c/T_h$ in the
following form

\begin{eqnarray}
P &\simeq& \frac{ 2\hbar\Gamma_h(\omega_h-\omega_c)(\omega_c - \tau\omega_h )}
 {3(\omega_c \gamma + \tau\omega_h )},\label{power6} 
 \\
P_\eta &\simeq& \frac{2\hbar \Gamma_h(\omega_c - \tau\omega_h) (\omega_h-\omega_c)^2}
 {3\omega_h(\omega_c \gamma + \tau\omega_h )}.\label{eco6}
\end{eqnarray}
%
%\begin{equation}
%P \simeq \frac{ 2\hbar\Gamma_h(\omega_h-\omega_c)(\omega_c - \tau\omega_h )}
% {3(\omega_c \gamma + \tau\omega_h )}, 
% \quad
%P_\eta \simeq \frac{2\hbar \Gamma_h(\omega_c - \tau\omega_h) (\omega_h-\omega_c)^2}
% {3\omega_h(\omega_c \gamma + \tau\omega_h )}.\label{eco6}
%\end{equation}
%%
\subsection*{Expressions for power at MEP and MP}
\subsubsection*{For fixed $\omega_h$}
Optimizing $P_\eta$ given in Eq. (\ref{eco6}) with respect to $\omega_c$ by setting
setting $\partial P/\partial\omega_c=0$, we have 
\begin{equation}
\omega^*_c = \frac{\gamma  \tau  \omega _h+\sqrt{\gamma +1} \sqrt{\tau } \sqrt{\gamma  \tau +8 \gamma +9 \tau } \omega _h-3 \tau  \omega _h}{4 \gamma }.
\label{eq1}
\end{equation}
Using Eq. (\ref{eq1}) in Eq. (\ref{power6}), we evaluate the expression for power at for the engine operating at 
MEP:
\begin{equation}
P^{P^*_\eta}_{\omega_h}=\frac{2\hbar\Gamma_h \omega_h\left(\gamma  (5 \tau +4)-3 \sqrt{\gamma +1} \sqrt{\tau } \sqrt{(\gamma +9) \tau +8 \gamma }+9 \tau \right)}{12\gamma ^2}. \label{peffp}
\end{equation}
Similarly, expression for optimal power is  given by
\begin{equation}
P^*_{\omega_h}=\frac{2\hbar\Gamma _h\omega_h\left(\gamma + 2\tau+\gamma\tau -2 \sqrt{(\gamma +1) \tau  (\gamma +\tau )} \right) }{3\gamma ^2}. \label{optimalp}
\end{equation}
\subsubsection*{For fixed $\omega_c$}
Since this case, Casus irreducibilis arises, in order to find the analytic expression for the efficiency, we have to take limits 
$\gamma\rightarrow 0$ and  $\gamma\rightarrow\infty$ in Eq. (\ref{EPhot}) before optimizing it. For 
$\gamma\rightarrow 0$, Eq. (\ref{EPhot}) is reduced to
\begin{equation}
P_{\eta(\gamma\rightarrow 0)} = \frac{2\hbar\Gamma_h(\omega_h-\omega_c)^2(\omega_c-\tau\omega_h)}{3\tau\omega_c\omega_h}.
\label{EPhot2}
\end{equation}
Keeping $\omega_c$ fixed, and optimizing with respect to $\omega_h$,  EMEP is evaluated at $ \eta_{_{YC}}$ value.
For $\gamma\rightarrow\infty$, Eq. (\ref{EPhot}) can be written as
\begin{equation}
P_{\eta(\gamma\rightarrow\infty)} = \frac{2\hbar\Gamma_c(\omega_h-\omega_c)^2(\omega_c-\tau\omega_h)}{3\omega_c\omega_h}.
\label{EPhot3}
\end{equation}
Optimization with respect to $\omega_h$ (fixed $\omega_c$) yields $\eta=\left(3-\sqrt{9-8\eta_C}\right)/2$.
%%
%\begin{equation}
%\eta = \frac{1}{2} \left(3-\sqrt{9-8\eta_C}\right).
%\end{equation}
%
%
\section{Casus Irreducibilis}
%In algebra, Casus irreducibilis arises while solving a cubic equation.
%The formal statement of the Casus irreducibilis 
%is that if a cubic polynomial is irreducible with rational coefficients 
%and has three real roots, then the roots of 
%the cubic equation are not expressible using real radicals and thus, 
%one must introduce expressions with complex radicals,
%even though the resulting expressions are actually real-valued.
%%
%It was proven by P. Wantzel in 1843 \cite{Kleiner2007}.  Using the 
%discriminant $D$ of the irreducible cubic equation, 
%one can decide whether the given equation is in Casus irreducibilis or not,
%via Cardano's formula \cite{Stewart1990}.
%
While solving a cubic equation, the case of Casus irreducibilis may arise \cite{Kleiner2007,Stewart1990}. 
Casus irreducibilis arises when 
the discriminant $D=18abcd-4b^3d+b^2c^2-4ac^3-27a^2d^2$ of a cubic equation 
\begin{equation}
a x^3 + b x^2 + c x + d  =0, \qquad (a, b, c, d \text{\, are real})
\end{equation}
is always positive, i.e., $D>0$.
In this case, all three roots of the cubic equation are real. If the roots cannot be found using the rational root test,
then the given polynomial is Casus irreducibilis and complex valued expressions are needed to 
express the roots in radicals.

In our case,  we have to solve the following cubic equation
\begin{equation}
 \tau ^2 \omega _h^3 + \tau  (2 \gamma +\tau ) \omega _c \omega _h^2 - (\gamma +2 \tau ) \omega _c^2 \omega _h
-
\gamma  \omega _c^3 =0. \label{Casus1}
\end{equation}
The discriminant $D$ of the above equation is given by
\begin{equation}
D = 4 (\gamma +1) \tau ^2 \left(8 \gamma ^3 \tau +\gamma ^3+12 \gamma ^2 \tau ^2+6 \gamma ^2 \tau +6 \gamma  \tau ^3+12 \gamma  \tau ^2+\tau ^4+8 \tau ^3\right) \omega _c^6.
\end{equation}
Since all the parameters $\omega_c, \gamma, \tau$ are positive, $D>0$. 
So the polynomial in Eq. (\ref{Casus1})
presents us with the case of Casus irreducibilis.
\label{casusir}
\end{widetext}

\bibliography{LaserQHE}
\bibliographystyle{apsrev4-1}

\end{document}